**Solitary and shock waves in discrete double power-law materials**

E.B. Herbold

*Department of Mechanical and Aerospace Engineering, University of California at San Diego, La Jolla, California 92093-0411, USA*

V.F. Nesterenko

*Department of Mechanical and Aerospace Engineering, University of California at San Diego, La Jolla, California 92093-0411, USA; Materials Science and Engineering Program, University of California at San Diego, La Jolla, California 92093-0418, USA*

A novel strongly nonlinear laminar metamaterial supporting new types of solitary and shock waves with impact energy mitigating capabilities is presented. It consists of steel plates with intermittent polymer toroidal rings acting as strongly nonlinear springs with large allowable strain. Their force-displacement relationship is described by the addition of two power-law relationships resulting in a solitary wave speed and width depending on the amplitude. This double nonlinearity allows splitting of an initial impulse into two separate strongly nonlinear solitary wave trains. Solitary and shock waves are observed experimentally and analyzed numerically in an assembly with Teflon o-rings.

PACS numbers: 05.45.Yv, 46.40.Cd, 43.25.+y, 45.70.-n

It is well known that strongly nonlinear compressive solitary waves exist in granular media composed of elastic spheres, which exhibit a Hertzian interaction at the contact point. The solitary wave solution of the long-wave approximation for discrete chains was discovered in [1] and reproduced in numerical analysis and experiments under various conditions and for different materials in [2-18].

The theoretical description of strongly nonlinear compressive solitary waves has mainly focused on power-law potentials between particles where the force-displacement relationship is described by $F \propto \delta^n$. Also, strongly nonlinear compression solitary waves were shown to exist for arbitrary "normal" nonlinear interaction between particles in the long wave approximation [2] and for a discrete chain [3]. A Hertzian system is an example of this type of strongly nonlinear material where the power-law exponent is $n$=3/2. The nonlinearity in these systems is due to the increase of surface area of the contact plane under compression. These materials can support pulses with amplitude less than critical value where plastic flow or fracture starts in the deformed area.

In this paper, a new form of a strongly nonlinear interaction between masses is introduced using toroidal o-rings. Here the geometrical nonlinearity under compression is caused by increase of width of an initial infinitesimally thin circular line of contact between toroidal o-rings and metal plates. In a first approximation, polymer or rubber o-rings can be considered massless nonlinear springs compared to the mass of rigid steel cylinders. The quasi-static deformation of o-rings has been experimentally investigated in [19-21]. The empirical form of the equation relating force to displacement is a double power-law where $F \propto \left(\delta^{3/2} + \delta^6\right)$. The unique structure of this relationship is represented by two regimes of nonlinear behavior where the Hertzian regime (first term) determines the dynamic behavior of small displacements ($\delta < 0.2$) and the higher order term dominates at larger deformations. The quasi-static force versus displacement relationship presented in [19, 21] is used as the basis for the equations of motion in the following numerical analysis,

$$\ddot{u}_i = A\left[(u_{i-1} - u_i)^{3/2} - (u_i - u_{i+1})^{3/2}\right] + B\left[(u_{i-1} - u_i)^6 - (u_i - u_{i+1})^6\right], \quad (1)$$

where $\delta_{i,i+1} = (u_i - u_{i+1})/d$, $u_i$ and $u_{i+1}$ denote the displacement of the cylinder on either sides of the o-ring from its equilibrium position, $A = 1.25\pi DE/md^{1/2}$ and $B = 50\pi DE/md^5$. The other constants in Eq. (1) are; $d$, which is the cross-sectional diameter of the toroid (o-ring), $D$ is the mean diameter of o-ring and $E$ is Young's modulus. Equation (1) describes experimental data reasonably well for relatively large strains $\delta \leq 0.45$ [19, 20]. The energy absorption of polymer or rubber per unit mass is significantly higher than in metals [19].

We neglect dissipation and gravitational effects in our numerical calculations in this investigation as they are negligible compared to the dynamic force. The behavior of the system can be qualitatively changed (tuned) with pre-compression with the application of an external force to the end particle in a chain of these elements [14,17,18].

A chain of 40 elements is used to numerically investigate the properties of this solitary wave at low and high impact velocities. Each element consists of a rigid cylinder and deformable o-rings. The properties of the Teflon o-rings are $E$=1.46 GPa [16,18,22], $D$=7.12 mm and $d$=1.76 mm and they are placed between cylinders with a mass of 3.276 g. The height, $h$, and diameter of the cylinders were 5 mm and 10 mm respectively. Impacting the chain at different velocities with a stainless steel sphere rapidly generated single solitary waves with different amplitudes if the mass of the sphere is less than the mass of the cylinder. The diameter and mass of the striker are 4.76 mm and 0.455g.

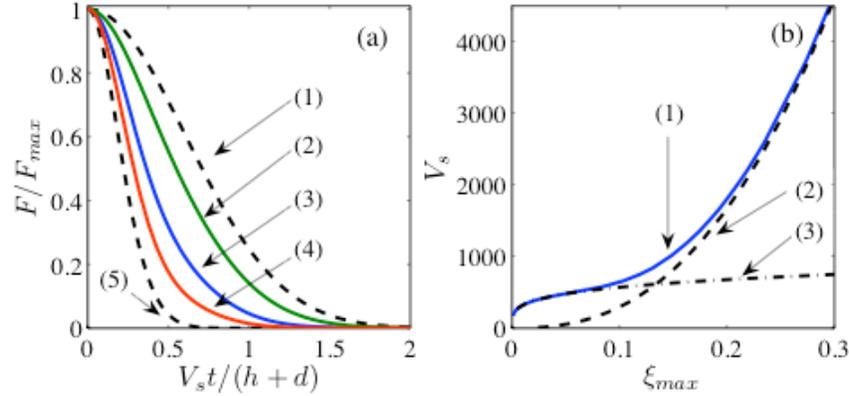

Figure 1: (a) Plot of normalized force vs. dimensionless time corresponding to different striker velocities $v_0$ in a solitary wave with 'sonic vacuum' initial conditions. Curve 1: $v_0$=1 m/s, only the Hertz law ($n$=3/2) is used in this case; curve 2: $v_0$=300 m/s; curve 3: $v_0$ =500 m/s; curve 4: $v_0$ =800 m/s; curve 5: $v_0$=1 m/s, only the higher order power-law relationship ($n$=6) is used in this case. (b) The dependence of solitary wave speed on the strain amplitude. Curve 1: double power-law interaction, Eq. (1); curve 2: power-law, $n$=6; curve 3: Hertz law, $n$=3/2.

The shape and duration of the pulse for different amplitudes are shown in Fig. 1 (a). In each of the five curves in Fig. 1(a) the normalized force between the 20[th] and 21[st] elements (where $F_{max}$ is the maximum force in the wave) is plotted versus nondimensional time representing half of the symmetric single solitary wave in a sonic vacuum. The value of $V_s$ used in the abscissa of Fig. 1(a) was based on the numerical results by dividing the distance between the 15[th] and 20[th] elements, 5($h$+$d$), by the time interval between arrival of the maximum strains at the corresponding points. The dotted lines in Fig. 1(a) show the shape of the pulse in two cases using only the exponent for the Hertzian interaction ($n=3/2$, curve 1) and the result corresponding to the case where only the second term in Eq. (1) was employed ($n=6$, curve 5). Intermediate cases when both parts of Eq. (1) are included and significant are shown in curves 2-4. In contrast to single power-law materials curves 2, 3 and 4 demonstrate that increasing solitary wave amplitude by increasing the velocities of the striker results in decreasing width down to about 3 elements in double power-law materials.

Figure 1(b) shows the dependence of wave speed on the maximum strain in the solitary wave for different interaction laws between cylinders. Each of the curves in Fig. 1(b) is the result of 26 calculations using 50 elements (cylinders and PTFE o-rings) for various striker impact velocities. Each of the simulations estimated the pulse speed and maximum strain in the pulse between the 25th and 30th particle. Curve 1 shows the pulse speed, $V_s$, dependence on maximum strain, $\xi_{max}$, for the double power-law materials. Curves 2 and 3 correspond to calculations using a single power-law with $n=6$ and $n=3/2$. Also, curves 2 and 3 have the power-law dependence identical to a long wave approximation [2] with a slight difference in the constant before the exponent. The agreement is better for $n=3/2$ than for $n=6$ due to a smaller width of the solitary wave, similar to the results observed in [9].

It is evident that the addition of the power-law term with exponent $n=6$ represents a qualitatively different case in comparison with Hertzian law (where $n=3/2$). It provides an opportunity for much stronger tunability of sound speed due to precompression and dependence of solitary wave speed on amplitude. This will also facilitate the splitting of initial pulse into two groups of solitary waves already very fast in Hertzian systems [1].

It is natural to expect that a larger exponent in a single power-law exponent will facilitate the splitting of an initial impulse into a train of solitary waves due to a shorter duration of solitary wave at larger exponents. It will also allow greater tunability of band gaps in two mass chains.

We observe that the wave speed converges to a single power-law relationship in the low and high regions of strain and that there is an intermediate region where the pulse speed is transitional. Note that in Fig. 1(b), curve 1 diverges from curve 3 into an intermediate region before it converges to curve 2. This suggests that one cannot additively apply the equations for $V_s$ using the corresponding exponents for a double power-law material to construct curve 1. However, as a consequence of the anharmonic approximation, an additive combination of two parts must be used to calculate the sound speed $c_0$ at some initial precompression,

$$c_0^2 = a^2 \left[ \frac{3}{2} A (a\xi_0)^{1/2} + 6B(a\xi_0)^5 \right], \qquad (5)$$

where $a=(d+h)$ and $h$ is the height of the cylinder.

Two numerical calculations were conducted to show how an initial impulse splits into a train of solitary waves in Hertzian and double power-law chains. A column of 400 elements (rigid cylinders and PTFE o-rings) was impacted by a striker with a mass equal to five times elements in the chain. In Fig. 2(a) the strain of the resulting wavetrain is shown for the case using the double power-law and Fig. 2(b) shows the case of a wavetrain in Hertzian chain. It is clear that the enveloping shape of the two wave trains shown in Fig. 2(a) and (b) are quite different. In Fig. 2(a) the amplitude of the pulses decreases gradually in contrast to the Hertzian system.

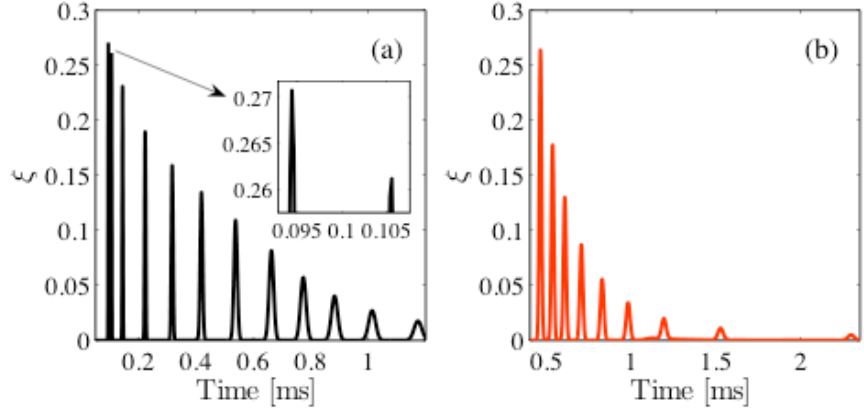

Figure 2: A train of solitary waves was created in a chain of 400 elements. The strain between the 50th and 51st cylinder is plotted as a function of time for a (a) double power-law system (Eq. (1)) with $v_0$=800 m/s and for a (b) Hertzian system with $v_0$=175 m/s ($n$=3/2). The inset in part (a) shows the amplitudes of the two leading waves.

In Table I the maximum strains and the wave speeds are given for the first nine pulses in double power-law materials and Hertzian systems shown in Fig. 2. In the double power-law system the velocity and maximum strain correspond to curve 1 in Fig. 1(b).

Table I: The amplitudes of the maximum strain and the speed of the first nine solitary waves shown in the Fig. 2(a) and (b).

| Pulse Number | Double Power-Law | | Hertzian | |
| --- | --- | --- | --- | --- |
| | $\xi_{max}$ | $V_s$ [m/s] | $\xi_{max}$ | $V_s$ [m/s] |
| 1 | 0.2707 | 3596 | 0.2638 | 726.8 |
| 2 | 0.2612 | 3380 | 0.1775 | 656.3 |
| 3 | 0.2306 | 2485 | 0.1298 | 609.0 |
| 4 | 0.1894 | 1610 | 0.0867 | 554.1 |
| 5 | 0.1588 | 1134 | 0.0554 | 493.4 |
| 6 | 0.1343 | 871.1 | 0.0344 | 436.1 |
| 7 | 0.1090 | 689.8 | 0.0197 | 400.0 |
| 8 | 0.0814 | 569.0 | 0.0110 | 334.7 |
| 9 | 0.0573 | 501.5 | 0.0050 | 271.5 |

Based on these values, the last wave in the train is completely in the Hertzian regime. The train of solitary waves in the Hertzian system shows that the pulse speed depends on the maximum strain in each wave corresponding to curve 3 in Fig. 1(b).

Experiments were performed to compare the strongly nonlinear dynamic behavior of a discrete system with numerical data. A system was assembled from 35 stainless steel cylinders and 34 PTFE o-rings (with a mass equal to $m=0.103$g) with properties similar to those used in numerical calculations. The aspect ratio of the o-ring diameters is $D/d=4.05$, which is within the experimental range investigated in [19-21] where the data are described well by Eq. (1) for o-rings with aspect ratios ranging from 3 to 30. The experimental setup is shown in Fig. 3.

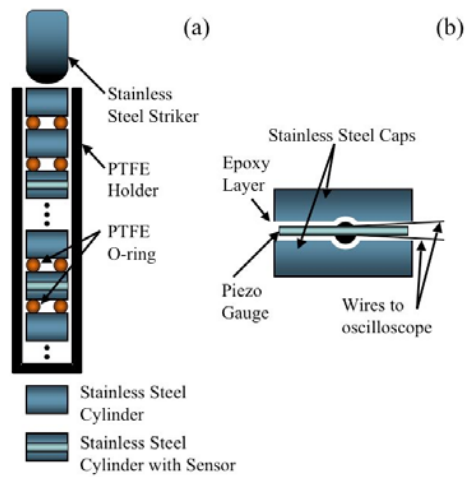

Figure 3: (a) Experimental setup of o-rings placed in between stainless steel cylinders and a striker with a hemispherical end. (b) A schematic diagram of the sensor embedded into the cylinder.

The column of cylinders and o-rings is placed in a hollow PTFE cylinder and two gauges composed of cylinders with imbedded piezo sensors are placed within the column in 3$^{rd}$ and 8$^{th}$ cylinders from the top (see Fig. 3(a)). Two sensors were placed using epoxy between two halves of a cylinder ensuring a similar mass and height to the other cylinders in the column (see Fig. 3 (b)). The positions of the sensors are the same for each experiment and corresponding numerical calculation presented here.

To create a single solitary wave the top of the assembly shown in Fig. 3 (a) was impacted with a stainless steel sphere whose diameter and mass were 4.76mm and 0.455g respectively. The initial velocity of the impacting sphere was $v_0$=2.1 m/s. The experimental and numerical results are shown in Fig. 4. It is remarkable that the amplitude of the pulses at the 3$^{rd}$ cylinder match so well without any fitting parameters. The wave amplitude and pulse speed correspond to the Hertzian regime. It is interesting that the quasistatic theory for point contact works well for the contact of toroids with planes. Also, it is clear that attenuation results in the area of compressed media behind the solitary wave as evident from Figure 4(a). This behavior may result in a complex two wave structure investigated in [23].

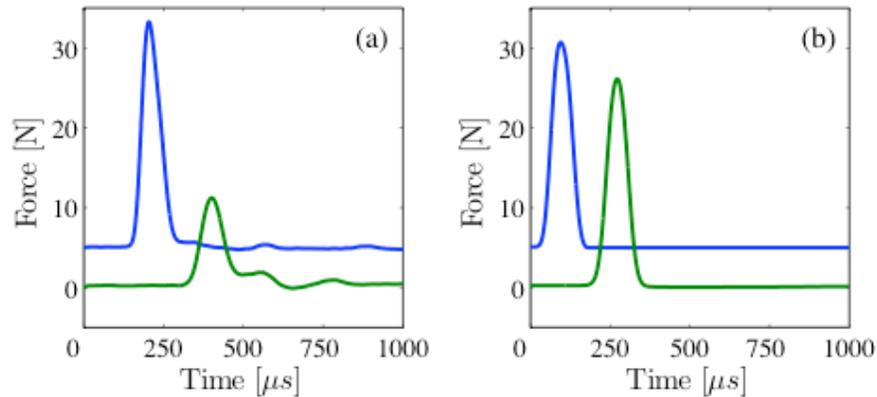

Figure 4: (a) Single pulse generated in experiment in a double power-law chain using PTFE o-rings; (b) numerical results corresponding to conditions of Fig. 4(a) using a 5 mm diameter steel striker (0.454g) with a velocity of $v_0$=2.1 m/s.

The average pulse speed between the two cylinders in Fig. 4(a) is 172 m/s and the second pulse is highly attenuated. The pulse speed in the calculations corresponding to Fig. 4(b) is 208 m/s, the difference is apparently due to the strong attenuation in experiments. To check if the pulse is stationary in numerical calculations an identical pulse to the one shown in Fig. 4(b) is created in a chain of 400 elements. The force amplitude of the wave at the $8^{th}$ cylinder was $F_8$=26.699N and the amplitude was $F_{50}$=26.693N at the $50^{th}$ particle. Such a small difference in force of these two pulses leads us to believe that we are looking at an almost fully developed stationary solitary wave at a very short distance from the end of the chain.

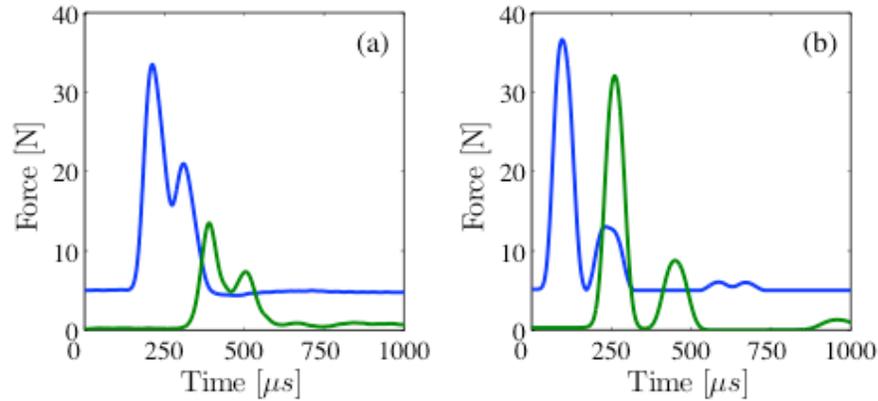

Figure 5: Two pulses generated in a double power-law chain assembled from stainless steel cylinders and Teflon o-rings. (a) Experimental results using a stainless steel impacting rod with a 10 mm radius of hemispherical tip with a velocity of $v_0$=0.44 m/s. (b) Numerical results using similar impact conditions.

In separate calculations we investigated the behavior of the collision of two identical solitary waves in double power-law materials. The leading solitary waves emerge with practically the same amplitudes after the collision followed by secondary solitary waves with amplitudes about $10^{-6}$ times smaller than the primary waves similar to [24].

To investigate if the presented assembly of cylinders and o-rings acts as strongly nonlinear system it is important to show that there is a tendency for a relatively long impulse to split into a train of solitary waves. To demonstrate this property we used a striker with mass $m$=6.236g equal to mass of two cylinders impacting a chain of 40 elements. The result is presented in Fig. 5. It is apparent from Fig. 5(a) that there is a tendency for two pulses to form in experiments, though attenuation prohibits the complete separation seen in numerical calculations presented in Fig. 5(b). Attenuation is known to prohibit the splitting of solitary waves in granular media in other cases as well [25] and qualitatively changes the profile of a shock wave [26] and the behavior of system under excitation by δ-function force [23]. Despite the difference in the splitting of the pulses the speeds of the leading pulses are reasonably close when comparing the experiments to the numerical calculations. The speed of the first pulse in experiment is 190 m/s (Fig. 5(a)) and 208 m/s in numerical calculations (Fig. 5(b)).

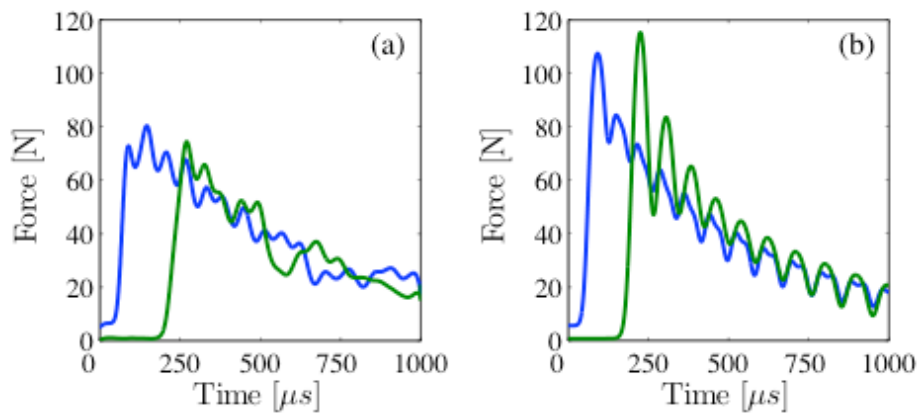

Figure 6: Oscillatory shock like waves in a chain assembled using PTFE o-rings and a stainless steel cylinder. (a) Experimental results, (b) Numerical results.

Shock like waves can be excited in this double power-law system using a relatively large mass of the striker. We used a striker with a mass equal to $m$=53.84g (15 times the mass of cylinders) and velocity $v_0$=0.767 m/s to investigate the type of propagating wave. It is especially important for this system because level of dissipation may dramatically change the shock profile from oscillatory to monotonic [26]. The experiments and numerical results are shown in Fig. 6(a) and (b), respectively. The amplitude of the pulse was within the range of validity for Hertz interaction law. The shock speed in experiments is $V_{sh}$=204 m/s and it is lower than 264 m/s in the calculations due to dissipation and resulting difference in shock amplitude (compare Fig. 6(a) and Fig. 6 (b)). The profile of shock wave is oscillatory, which means that the effective viscosity (if dissipation is linearly dependent on relative velocities of cylinders) is below critical value equal to $mV_{sh}/2a$, where $a=(h+d)$ is the distance between centers of cylinders [26]. In our case a value of critical viscosity is about 50 Ns/m.

In conclusion we emphasize that a novel strongly nonlinear laminate material was proposed and experimentally realized. Toroidal o-rings were used as strongly nonlinear springs that introduce a double power-law for the interaction between rigid masses. Large strains are allowed for polymer (or rubber) o-rings that may result in useful impact mitigating capabilities. Numerical calculations and experimental results demonstrated the existence of novel solitary waves in this material, which are qualitatively different from the discrete system with a single power-law interaction force. The shape of the solitary wave in the former case depends on the amplitude and they are highly tunable. We observed a qualitatively new behavior of splitting of initial pulse into a train of solitary waves in two separate nonlinear regimes in numerical calculations, and oscillatory shock waves in experiments. Despite the dissipation in experiments, the numerical calculations give a reasonable estimate of the pulse speed based on quasi-static characteristics of o-rings.

The authors wish to acknowledge the support of this work by the U.S. NSF (Grant No. DCMS03013220).

[1] V.F. Nesterenko, Prikl. Mekh. Tekh. Fiz. 24, 136 (1983) [J. Appl. Mech. Tech. Phys. 24, 733 (1984)].

[2] V.F. Nesterenko, Dynamics of Heterogeneous Materials (Springer-Verlag, New York, 2001).

[3] G. Friesecke and J.A.D. Wattis, Commun. Math. Phys. 161, 391 (1994).

[4] A. Chatterjee, Physical Review E 59, 5912 (1999).

[5] E.J. Hinch and S. Saint-Jean, Proc. R. Soc. Lond. A 455, 3201 (1999).

[6] J. Hong, J-Y Ji, H. Kim, Phys. Rev. Lett. 82, 3058 (1999).

[7] E. Hascoet, and H.J. Herrmann, Eur. Phys. J. B 14, 183 (2000).

[8] E. Hascoet, and E.J. Hinch, Phys. Rev. E 66, 011307 (2002).

[9] A. Rosas, K. Lindenberg, Physical Review E 69, 037601 (2004).

[10] A. Sokolow, E.G. Bittle and S. Sen, Eur. Phys. Lett. 77, 24002 (2007).

[11] A.N. Lazaridi and V.F. Nesterenko, Prikl. Mekh. Fiz. 26, 115 (1985) [J. Appl. Mech. Tech. Phys. 26, 405 (1985)].

[12] A. Shukla, M.H. Sadd, Y. Xu, and Q.M. Tai, J. Mech. Phys. Solids 41, 1795 (1993).

[13] Y. Zhu, F. Sienkiewicz, A. Shukla, and M. Sadd, J. Eng. Mech. 10, 1050 (1997).

[14] C. Coste, E. Falcon and S. Fauve, Physical Review E 56, 6104 (1997).

[15] S. Job, F. Melo, S. Sen, and A. Sokolow, Phys. Rev. Lett. 94, 0178002 (2005).

[16] C. Daraio, V.F. Nesterenko, E. Herbold, and S. Jin, Physical Review E, 72, 016603 (2005).

[17] C. Daraio, V.F. Nesterenko, Physical Review E, 73, 026612 (2006).

[18] C. Daraio, V.F. Nesterenko, E.B. Herbold, S. Jin, Physical Review E, 73, 026610 (2006).

[19] P.K. Freakley and A.R. Payne, Theory and Practice of Engineering with Rubber (Applied Science, London, 1978).

[20] P.B. Lindley, J. Strain Anal. 1, 190 (1966).

[21] P.B. Lindley, Use of Rubber in Engineering Conf. Proc. at Imp. College of Sci. and Tech., London, Ch. 1 and 2(1966). (Edited by P.W. Allen, P.B. Lindley and A.R. Payne).

[22] W.J. Carter and S.P. Marsh, Los Alamos National Laboratory Report No. LA-13006-MS, 1995 (unpublished).

[23] A. Rosas, A.H. Romero, V.F. Nesterenko, and K. Lindenberg, Phys. Rev. Lett. 98, 164301 (2007)

[24] M. Manciu, S. Sen, and A.J. Hurd, Physical Review E, 63, 016614 (2000).

[25] E.B. Herbold, V.F. Nesterenko, and C. Daraio, *in Proceedings of the Conference of the American Physical Society Topical Group on Shock Compression of Condensed Matter*, AIP Conf. Proc. No. 845 (AIP, Melville, NY, 2006), pp. 1523-1526; e-print cond-mat/0512367.

[26] E.B. Herbold and V.F. Nesterenko, Phys. Rev. E 75, 021304 (2007).